\documentclass[journal,letterpaper]{IEEEtran}
\usepackage{cite}      
\usepackage{graphicx}  
\usepackage{amsmath}   
\usepackage{url}
\usepackage{array}
\usepackage{multirow}
\usepackage{pstricks}
\usepackage{pst-node}
\usepackage{booktabs}
\usepackage{balance}
\usepackage{color}

\makeatletter
\let\NAT@parse\undefined
\makeatother

\newcommand{\etal}{{\it et al.}}
\begin{document}

\title{Open Material Property Library With Native Simulation Tool Integrations -- MASTO}

\author{Antti Stenvall and Valtteri Lahtinen%
\thanks{Manuscript received August 28, 2017.}
\thanks{The authors are with Tampere University of Technology, Laboratory of Electrical Energy Engineering, Group of Modelling and Superconductivity, P.O. Box 692, FIN-33101 Tampere, Finland (phone: +385-40-8490403; fax: +358-3-31152160; e-mail: antti.stenvall@tut.fi; www: http://www.tut.fi).}
\thanks{The authors would like to acknowledge the following financial support: Academy of Finland project number 287027.}
}

\markboth{}{}

\maketitle

\begin{abstract}
Reliable material property data is crucial for trustworthy simulations throughout different areas of engineering.
Special care must be taken when materials at extreme conditions are under study.
Superconductors and devices assembled from superconductors and other materials, like superconducting magnets, are often operated at such extreme conditions: at low temperatures under high magnetic fields and stresses.
Typically, some library or database is used for getting the data.
We have started to develop a database for storing all kind of material property data online called Open Material Property Library With Native Simulation Tool Integrations -- MASTO.
The data that can be imported includes, but is not limited to, anisotropic critical current surfaces for high temperature superconducting materials, electrical resistivities as a function of temperature, RRR and magnetic field, general fits for describing material behaviour etc. 
Data can also depend on other data and it can be versioned to guarantee permanent access.
The guiding idea in MASTO is to build easy-to-use integration for various programming languages, modelling frameworks and simulation software. 
Currently, a full-fledged integration is built for MATLAB to allow users to fetch and use data with one-liners.
In this paper we briefly review some of the material property databases commonly used in superconductor modelling, present a case study showing how selection of the material property data can influence the simulation results, and introduce the principal ideas behind MASTO.
This work serves as the reference document for citing MASTO when it is used in simulations.
\end{abstract}

\begin{IEEEkeywords}
experimental data, material property database, numerical modelling, simulations
\end{IEEEkeywords}

\IEEEpeerreviewmaketitle

\section{Introduction}
\IEEEPARstart{I}{n} all simulation work a guiding principle is {\it garbage in means garbage out}. 
The inputs of typical simulations include the device under study, the operation conditions and the material properties characterizing the components of the device.
The model of the device under study can typically be taken, and often simplified, from the design drawings, or from the constructions.
Operation conditions may be known or certain conditions are sought.
It can be difficult to find reliable material property data for special materials or typical materials at extreme operation conditions.
Characterization is often possible but can be very time consuming.
Therefore, modellers often rely on known sources -- material property libraries or databases.
In principle they are the same thing with different names.

Three larger material property libraries, databases or sources are well-known in the superconducting magnet community, among others.
The data from the Cyogenics Technologies Group at the Material Measurement Laboratory, National Institute of Standards and Technology (USA) is often called NIST data in modellers' jargon~\cite{nist}.
It includes thermal conductivities, electrical resistivities and specific heats, among other properties, for materials commonly used in constructing superconducting magnets such as different aluminum alloys, stainless steels and copper. 
Fits as a function of temperature, and in some cases for different material purities and as a function of magnetic flux density, are provided with the data.
The data are available via web pages of NIST in HTML format from which one can copy the fits and the coefficients to one's simulation tool.
The data are originally from measurements at NIST.
The NIST database represents data from a national level public organization.

MATPRO -- A Computer Library of Material Property at Cryogenic Temperature represents another such library~\cite{matpro}.
MATPRO was developed in a collaboration between CERN and University of Milano, Italy. 
Typical materials (insulators, metals, alloys) utilized in constructing superconducting magnets are included in MATPRO including the most common low and high temperature superconductors.
The data in MATPRO are restricted to densities, specific heats, electrical resistivities and thermal conductivities of materials.
In the electrical resistivity and thermal conductivity data, the magnetic field and RRR dependence are also included in addition to the temperature dependence. 
MATPRO is written in Fortran77 and it also has a Windows compatible executable that one can use via the command prompt.
The executable can be used to extract numerical data in ASCII format.
The data in MATPRO are collected from different sources, all documented in~\cite{matpro}.

CryoComp is a commercial material property library from Eckels Engineering Inc.~\cite{cryocomp}.
A Windows compatible executable is provided for a license buyer.
Data can be accessed via an index and can be represented with tables, which can be directly saved or copied and pasted into, for example, Excel.
This database also includes typical materials for superconducting magnet design which are grouped by type in the index. 
For example, seven different nickel alloys are included.
One can also add private data into the database and use it via the same graphical user interface as the built-in data.
Data in CryoComp comes without references and without any warranty or even a guarantee of suitability for the condition for which it is provided. 
One should, however, note that even though this may sound strange, this is standard license agreement text.

Common to the presented three material property libraries is that they are standalone ones, not integrated to modelling software, and not extendable by people not working directly with these software.
Data from the presented databases, as well as others, has been partially collected and implemented into some special software, like for performing quench analyses in ROXIE~\cite{roxie}.
This integration, however, is not straightforward as it is merely an export, and requires implementation of the properties to the software, rather than a dependency which updates itself when the original software is updated.

This paper introduces an ongoing effort to build a material property database with the same principles as social coding networks like GitHub~\cite{github} and tools utilizing shared software like Composer~\cite{composer} or Bower~\cite{bower}, but with the possibility of features following scientific practices.
This includes optional data review to promote their reliability.
The database is called The Open Material Property Library With Native Simulation Tool Integrations -- MASTO.
A domain name (\url{masto.eu.com}) for MASTO has been reserved and an early prototype is accessible there via a web browser.
MASTO will not be limited to just hosting for material property data, but also for networking among all the different stakeholders around materials and their properties. 
The completely new way of acting with material property data that MASTO represents aims to establish a sustainable, open, extendable and reliable material property network from which one can find materials and their properties, find experts to characterize materials and which one can use as a store window to make materials available for possible customers.

In this paper we first compare selected data from the three introduced material property databases -- NIST, MATPRO and CryoComp -- and study how the variation in the data influences on modelling results. 
The aim of this comparison is not to question the reliability of the databases, but to emphasize how important it is to select the material properties for simulations and how to get error estimates for the results by using different data sources.
After this we introduce the fundamental ideas in MASTO and present its MATLAB integration.
Finally, conclusions are drawn.

\section{Influence of Material Property Data on Simulation Results -- a Case Study}

A simple way to estimate hot spot temperature in a superconducting magnet follows the so-called adiabatic MIITs~\cite{todesco-2013} approach in which the heat conduction is neglected and the hot spot temperature can be directly estimated from the current decay curve -- either measured or simulated. 
The adiabatic heat balance equation:
\begin{equation}
c\frac{\partial T}{\partial t} = \rho \left(\frac{I}{A}\right)^2,
\end{equation}
where $c$ is the volumetric effective specific heat, $\rho$ is the stabilizer resistivity, $I$ is the current and $A$ is the cross-sectional area of the stabilizer, can be separated into material properties and current-dependent terms, and integrated separately as:
\begin{equation}
\label{equ:miits}
\int_{T_{op}}^{T_{max}}\frac{A^2c}{\rho}{\rm d}T = \int_0^\infty I^2{\rm d}t,
\end{equation}
where $T_{op}$ corresponds to the operation temperature at which heat starts to be generated and $T_{max}$ corresponds to the upper limit of the hot spot temperature, typically at the origin of the quench, at which the integrals on the right and left hand sides are equal. 
For the material properties, those averaged over the magnet's unit cell are used.
Therefore, by measuring the current decay curve in a quench experiment, which is straightforward to measure, one can find $T_{max}$. 
One should note that because the heat conduction, and any cooling, is neglected, $T_{max}$ does not represent the hot spot temperature, but an upper limit for it -- within the error of material property data and assumption of isothermal unit cells.
MIITs abbreviation comes from the scaled units: mega, current, current and time and~(\ref{equ:miits}) is called the MIITs equation.

The computation of the material property integral in the MIITs equation to different values of $T_{max}$ can reveal the sensitivity of the estimated hot spot temperature to changes in material properties.
To study this, we use the three presented material property sources: NIST, MATPRO and CryoComp.
To simplify, we use only material properties of copper. 
We consider RRR of 100 and magnetic flux densities of 0~T and 12~T to take into account also the magnetoresistivity which at low temperatures is significant.

The databases use different definitions for RRR.
Here, we define RRR as the ratio $\rho(273~{\rm K})/\rho(4.2~{\rm K})$ at 0~T.
The RRR values for the ratio of 100 defined accordingly corresponding to the RRR values in the different databases are the following: NIST 99.7, MATPRO 99 and CryoComp 100.  

We tabulate the material property values from 4 to 300~K with a spacing of 1~K and perform the integration with the trapezoidal method.
Our $T_{op}$ is 4.2~K.
We normalize our results to the NIST data.
Because the material properties change by orders of magnitude as a function of temperature, the differences are better visible in the normalized graphs.

\begin{figure}[ht!]  
	\centering
	\includegraphics[width=6.9cm]{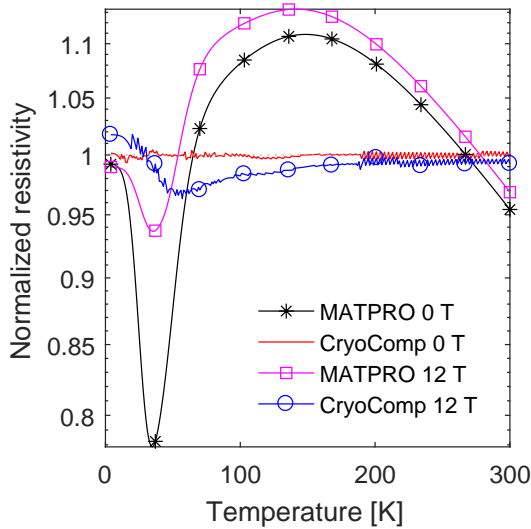}
	\caption{\label{fig1} Normalized copper resistivities from the MATPRO and CryoMat databases for RRR=100 and 0~T and 12~T. Normalization is done according to the corresponding data from NIST database.}
\end{figure}

Fig.~\ref{fig1} compares the resistivities from the three different databases. 
As can be seen, CryoComp data at 0 T is almost the same as NIST data, differing by less than 0.5\%. 
At 12~T the largest difference is still less than 4\%.
This characterizes the difficulty of describing resisitivity as a function of RRR and magnetic flux density.
NIST and MATPRO data differ substantially both at low temperatures and intermediate temperatures. 
At 0~T field, the MATPRO data plummets to 22\% below the NIST value around 35~K.
At 12~T this peak is reduced to 6\% at the same temperature.
Another local maximum in relative difference between the MATPRO and NIST data can be found at 147~K and 142~K for 0~T and 12~T, respectively.
In the case of 12~T, the difference is 13\%.

\begin{figure}[ht!]
    \vspace*{2mm}
	\centering
	\includegraphics[width=6.9cm]{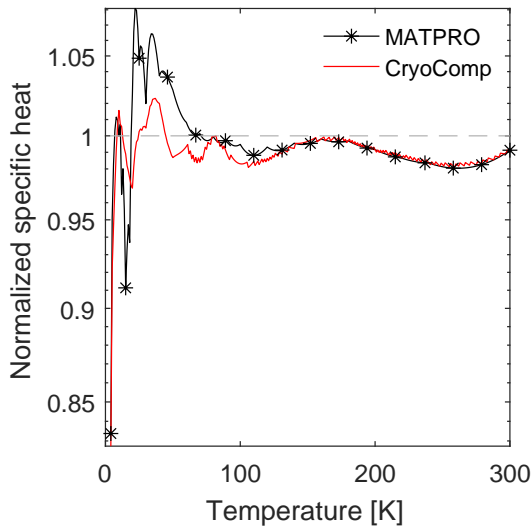}
	\caption{\label{fig2} Normalized specific heat of copper from MATPRO and CryoMat databases. Normalization is done to the NIST data.}
\end{figure}

Fig.~\ref{fig2} displays the difference in the values of specific heat. 
At low temperatures the relative variation oscillates whereas at higher temperatures it changes less rapidly. 
Above 160~K CryoComp and MATPRO data fit very well together and differ at most 2\% from the NIST data.
One should keep in mind that a small absolute variation at low temperature results in a larger variation in relative error than at high temperatures.
The ratio of copper's specific heat at 300~K to 4.2~K is more than 4000.

To study how these variations in material property values influence modelling, it is important to consider their integrals. 
For example, the typical quantity of interest in quench modelling is the hot spot temperature at the end of quench.
Because the ratio of the specific heat and the resistivity play a crucial role in the temperature increase, the integral of that is the most visible parameter showing the difference.
Further, variation of this at low temperatures can have an influence on predicting, for example the quench protection system delay.
Such sensitivity analysis has been done in~\cite{salmi-2014}.

\begin{figure}[t!]
	\includegraphics[width=6.9cm]{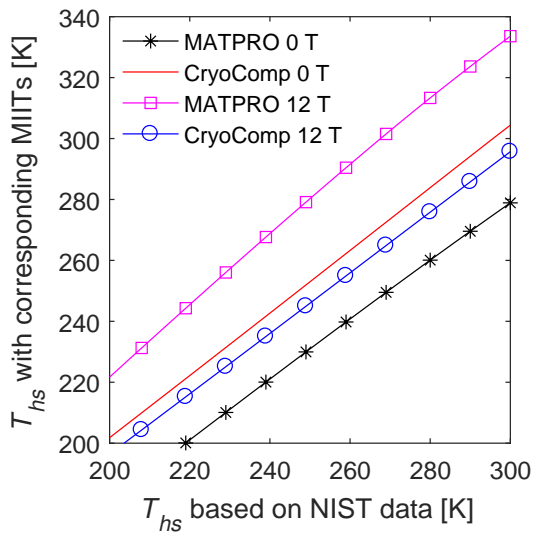}
	\caption{\label{fig3} $T_{hs}$ predicted by material property data from MATPRO and CryoMat databases as a function of $T_{hs}$ predicted by the data from NIST database at 0~T and 12~T, when the operation temperature was 4.2~K.}
\end{figure}

Fig.~\ref{fig3} displays how different current decay curves, i.e. MIITs, predict different hot spot temperatures when material properties are taken from different libraries. 
We do not present actual MIITs values but only compare how the predicted $T_{hs}$ changes. 
At 0~T, MATPRO is more optimistic than NIST which is still slightly more optimistic than CryoComp.
If the MIITs correspond to such a value that NIST data gives 300~K for $T_{hs}$, the MATPRO data predicts $T_{hs}$ of only 279~K.
The corresponding value for CryoComp data is 304~K.
In particular, the very low resistivity of MATPRO below 60~K influences its considerably lower value.
Interestingly, at 12 T the situation is reversed.
With the MIITs corresponding to NIST data of 300~K, MATPRO predicts 334~K and CryoComp 296~K.
Because as a function of field both of the datasets intersect NIST data, a field value which matches at a given temperature for NIST and MATPRO or NIST and CryoComp can be found.
However, real situations are seldom like that, usually the field changes during the current decay and is different in different parts of the system.

There exists large variation in the literature data for material properties of even the most common material, and possibly the most important, from the stability point of view, copper.
This variation can have a significant influence when deciding if devices designed on their limits are feasible or not.
Further, to include this data into a simple simulation tool, we needed to implement relevant functions to our simulation tool manually by extracting the data from the databases and compiling to an appropriate format.
In the case of a database with simulation tool integrations, it would be easy to change the data source and re-run the simulations to get an estimate for the reliability of the results.

\section{Principal ideas behind MASTO and its MATLAB integration}

MASTO is a new kind of effort for constructing a centralized database for material property data that is also integrated to simulation tools.
The level of integration will depend on the particular tool at hand.

Critical questions in a centralized effort for building a material property database are
\begin{itemize}
	\item How to ensure the reliability of data?
	\item How to ensure that no user is blocking another one from entering similar data?
	\item How to ensure the persistence of data?
\end{itemize}

In a fully open system aimed at distributing information, such as arXiv~\cite{arxiv}, GitHub or any social media platform like Twitter and Instagram, in principle anyone can make anything visible for all web users. 
This is contrary to the scientific peer reviewing practice, in which experts check the submitted material beforehand and communicate their findings to editors who decide if revision is needed, material can be accepted or if it must be rejected.
Still, arXiv is popular among scientists, people upload a lot of software to GitHub and social media is used as the main information source by many people.
To increase the reliability of data, MASTO features an optional {\it peer review} that one submitting a material can ask for. 
In this case, the foreseen MASTO editorial and development office finds experts to blind review the data.
One can use the advanced search functionality to only search among the {\it reviewed} materials.
Other options for one browsing MASTO in estimating the reliability of data include a {\it star ranking} from 1 to 5 (similar to systems for evaluating e.g. movies) or searching for materials marked as {\it confirmed}.
Naturally, all the uploaded data can include citations to literature and description of the characterization mentioning the standards that have been followed.

Stainless steel, for example, means different things to different people; and even a standard such as  AISI 316L could be in question. 
The microstructure and manufacturing history has a non-negligible effect on the yield strength.
Therefore, a stainless steel expert must be able to identify her steels in detail while a typical user can import data that is named just as stainless steel.
This leads to the requirement of non-blocking namespaces for materials.
To allow this, materials are organized under {\it communities}.
The name of each community is unique in the system, but different communities can have {\it materials} with the same names.
A material is an entity in which a user has composed her data in a way she wants to.
For example, a material can be the yield strength of a stainless steel having a certain manufacturing history, or it can be a list of densities of all solid materials at standard temperature and pressure conditions (STP)~\cite{iupac,iupac-online}.
The communities are the data sources that the editorial and development office can tag as trusted in a similar way as in Twitter one can identify that the account \texttt{@realDonaldTrump} really belongs to the 45$^{\rm th}$ president of the United States and is not one of the many parody accounts.
The administration of communities can be distributed and therefore many researchers can contribute to the same community.
A unique identifier for a material property, data, fit etc, can be \texttt{experts/aisi-304}, with format \texttt{community/material}.
This community idea is also in use in GitHub for sharing code.
Communities can also have private material properties for internal use.

The International Digital Object Identifier Foundation~\cite{doi} has developed a wide spread way to identify publications and their locations in the internet allowing publishers to change the actual links in case of content management system upgrades etc.
In the MASTO system there is an additional dimension for this, as material property data can have various {\it versions}.
Consider one entering electrical resistivity of copper as a function of temperature there.
Because some code can depend on this, this cannot be removed from the system any more, in the same way as one cannot undo a published publication.
Therefore, when the data owner updates this material to also include magnetic field dependence and the effect of RRR, a new version is created from the data.
The versioning system follows the schema $x.y.z$, where $z$ means a small fix, or a bug fix, e.g. in data or its description; $y$ means minor improvement, or addition of new simulation software dependency, to the data that kees the same interface; and $x$ is a major upgrade.
Because the guiding idea is to make MASTO compatible with various simulation software, one can detail the version to use.
Therefore, persistence means identifying data like \texttt{experts/aisi-304/1.0.5} or in a simulation tool to have a dependency
\texttt{experts/aisi-304/1.\^}, which means the latest with major version 1.
This allows one to always find the data from MASTO, no matter if the internal linking system is changed.

MASTO offers a Representational State Transfer (REST) architecture~\cite{fielding-2000} based application programming interface to fetch, import and update data via HTTP. 
Currently, an utility package is built to MATLAB to allow full MASTO integration to in-house software~\cite{masto-matlab}, but the development of other integrations is underway.
The utility package can be found from MASTO community \texttt{stenvala}\cite{masto-stenvala} and package \texttt{utils}.
With this utility package (that an initialization script will install), for example densities of basic elements can be fetched with command \texttt{masto.stenvala.utils.latest.require( 'stenvala',  'element-densities');}.
To get the density of solid copper at STP conditions, one uses command
\texttt{masto.stenvala.elementDensities.latest( 'cu', 'solid')}.

\section{Conclusions}

Several material property databases or libraries for materials utilized in cryogenic environments exist.
Some of these are openly available on the internet, some meet the definition of properietary software and some are developed in collaboration between research institutes.
Typically, common to all of these are that when material property data for a given material at given conditions are fetched, it may differ, people cannot contribute to the libraries and any software integration must be done manually.

We studied three different databases called NIST, MATPRO and CryoComp and considered the electrical resistivity and specific heat of copper.
We considered a simple adiabatic Joule heating case and showed that at 0~T when NIST data predicted a temperature increase from 4.2~K to 300~K, MATPRO data gave only 279~K. With CryoComp data the result was 304~K. 
At 12~T the corresponding numbers were 334~K for MATPRO and 296~K for CryoComp.
We presented an ongoing effort to construct a new online material property database, 
Open Material Property Library With Native Simulation Tool Integrations -- MASTO, to which anyone can contribute with one's own material property data, the credibility of data can be assessed in multiple ways, where persistence of data as well as its versioning is guaranteed, and the material properties can be linked as dependencies to external software with no programming effort.
We aim to make MASTO a new research infrastructure connecting different people: modellers, experimentalists, material providers etc. around material property data.

\balance

\end{document}